# Circumventing cracking in grading 316L stainless steel to Monel400 through compositional modifications

Zhening Yang[a], Alexander Richter[a], Hui Sun[a], Zi-Kui Liu[a], Allison M. Beese[a, b*]

[a] Department of Materials Science and Engineering, Pennsylvania State University, University Park, PA 16802, USA

[b] Department of Mechanical Engineering, The Pennsylvania State University, University Park, PA 16802, USA

* Corresponding author: beese@matse.psu.edu

**Abstract**

In joining Fe-alloys and Cu-containing alloys to access the high strength of steels and corrosion resistance of Cu-alloy, cracking is widely observed due to the significant Cu microsegregation during the solidification process, resulting in an interdendritic Cu-rich liquid film at the end of solidification. By fabricating functionally graded materials (FGMs) that incorporate additional elements like Ni in the transition region between these terminal alloy classes, the hot cracking can be reduced. In the present work, the joining of stainless steel 316L (SS316L) and Monel400 by modifying the Ni concentration in the gradient region was studied. A new hot cracking criterion based on hybrid Scheil-equilibrium approach was developed and validated with monolithic multi-layer samples within the SS316L-Ni-Monel400 three-alloy system and an SS316L to 55/45 wt% SS316L/Ni to Monel400 FGM sample fabricated by direct energy deposition (DED) process. The new hot cracking criterion, based on the hybrid Scheil-equilibrium approach, is expected to help design FGM paths between other Fe-alloys and Cu-containing alloys as well.

**Keywords:**

Functionally graded material; CALPHAD modeling; hot cracking; crack susceptibility; additive manufacturing

# 1. Introduction

The alloy Monel400 is widely used in nuclear power plant and naval applications due to its high corrosion resistance and ductility [1]. Monel400 has been welded with stainless steel to take advantage of the high corrosion resistance of Monel and the structural support provided by the more cost-effective stainless steel [2]–[4]. However, welding Fe-based alloys to Cu-based alloys can result in cracking as Cu is segregated into the liquid during the solidification process, resulting in solidification or liquation cracking during fabrication. Studies on cracking when joining Cu-containing alloys and Fe-alloys have been reported: Dharmendra et al. deposited nickel aluminum bronze (NAB) on top of stainless steel 316L (SS316L) and observed cracking in the heat affected zone (HAZ) at the SS316L/NAB interface [5], which was attributed to the presence of Cu-rich liquid in the intergranular regions within SS316L; Liquation cracking has also been reported when welding stainless steel 304 (SS304) and T2 copper by Li et al. [6].

Functionally graded materials (FGMs) with spatially varying compositions can be fabricated using directed energy deposition (DED) additive manufacturing (AM) by changing the relative flow rates of powder feedstocks into the melt pool during fabrication [7]–[9]. As cracking due to intermetallic phases or hot cracking may occur when mixing dissimilar alloys, researchers have investigated the design of compositional paths between terminal alloys in FGMs to avoid cracking in their joining [10][11].

One method for circumventing cracking involves process modification to adjust the cooling rates during fabrication to mitigate Cu microsegregation during solidification. However, this requires modifications to the fabrication equipment, such as incorporating a heating system. Another method to avoid cracking when joining Fe- and Cu-containing alloys is to introduce another element or alloy that reduces Cu microsegregation during solidification process, e.g., a Ni-based intermediate alloy may be introduced between to take advantage of the high solubility between Ni and Cu. Successful joining of Fe-alloys and Cu-containing alloys with a Ni-based alloy intermediate layer has been reported in the literature for both welding and AM processes [2], [4], [12]–[15]. In welding, ERNICrFe-3 electrodes have been used as the filler material when welding SS316 and Monel alloys [2][4]. In DED AM, Ni-rich alloy Deloro 22 has been used as an intermediate alloy through which to join SS316L to Cu in ref. [12] and to join H13 tool steel to Cu in ref. [13]. Additionally, Inconel 718 (IN718) has been used as an intermediate alloy through which to join SS316L and Cu through DED AM in ref. [14]-[15]. Noecker et al. studied solidification cracking in the Cu-Fe-Ni ternary system by experimentally depositing Cu-Fe-Ni alloys using the gas tungsten arc weld (GTAW) process and studied the relationship between cracking and Scheil calculations results [16][17]. The cracking in the Cu-Fe-Ni ternary system was found to be associated with the solidification temperature range, amount of terminal liquid, and liquid distribution during the solidification process in ref. [17]. However, the study focused on the Cr-Fe-Ni ternary system and did not establish a quantitative relationship between the cracking and Scheil results.

In the present study, the variation in composition at the solid-liquid interface as a function of solid fraction during solidification were taken into account to avoid cracking when designing FGM compositional paths in alloy systems containing Cu, Fe, and Ni, employing the previously developed hybrid Scheil-equilibrium approach, where equilibrium simulations were carried out at a specific temperature for the compositions during the solidification process from the Scheil simulation to predict phase amount under the solidus temperature considering solidification microsegregation [18]. The severe Cu microsegregation resulting in a significant composition difference between the dendrite-core and interdendritic region at the end of the solidification process was identified as the reason for the cracking in the SS316L-Ni-Monel400 system. With the severe Cu microsegregation, the hybrid Scheil-equilibrium simulations will predict the formation of a secondary Cu-rich FCC, which can be taken as an indicator of sufficiently drastic Cu microsegregation leading to cracking. Thus, a new hot cracking criterion for the SS316L-Ni-Monel400 system based on the hybrid Scheil-equilibrium approach was developed. Consequently, multi-layer samples over a wide range of compositions in the SS316L-Ni-Monel400 three-alloy system were deposited using DED AM and characterized for cracking and Cu microsegregation. The hybrid Scheil-equilibrium method successfully predicted cracking in these single-composition samples. Additionally, an FGM sample grading from SS316L to 55/45 wt% SS316L/Ni to Monel400 was designed, fabricated, and characterized. The experimental results from this sample were used to verify the cracking map calculated using the hybrid Scheil-equilibrium approach, which provides a means for designing FGM compositional pathways between other Fe-alloys and Cu-containing alloys.

## 2. Experimental methods

### 2.1 Hybrid Scheil-equilibrium approach to predict formation of Cu-rich FCC phase

The cracking in the SS316L-Ni-Monel400 three-alloy system is due to Cu microsegregation to interdendritic regions during the solidification process. The difference in melting temperature between the solid and remaining liquid at the end of solidification process, combined with the infiltration of the remaining liquid between dendrites, due to the small wetting angle between the Cu-rich liquid and the Fe-rich dendrites [5], leads to the formation of interdendritic liquid films at the end of the solidification process or within the heat-affected zone (HAZ) during subsequent layer deposition, ultimately resulting in solidification cracking or liquation cracking.

**Figure 1** shows the Scheil results for three compositions (Samples are discussed in Sections 3.1 and 3.3). While the 70/30 wt% SS316L/Monel400 sample was not fabricated, it is shown here to provide three example compositions with the same amount of Cu concentration. Cracks were observed in the 40 wt% and 60 wt% Monel400 samples (balance SS316L). It can be seen that for the compositions with cracking, there is a wide temperature range at the end of the solidification process with no solid fraction increase (see **Figure 1(a)**), indicating that there is time for the remaining liquid to flow and infiltrate between the dendrites to form liquid films. The reason for the drastic temperature drop at the end of the solidification process is shown in **Figure 1(b)** by the drastic increase in Cu concentration at the end of the solidification process, indicating that the remaining interdendritic liquid in **Figure 1(a)** is almost pure Cu. **Figure 1** also shows that with increased Ni concentration, the solid fraction change as a function of temperature (**Figure 1(a)**) and the composition change as a function of solid fraction (**Figure 1(b)**) are more gradual, thus resulting in reduced crack susceptibility.

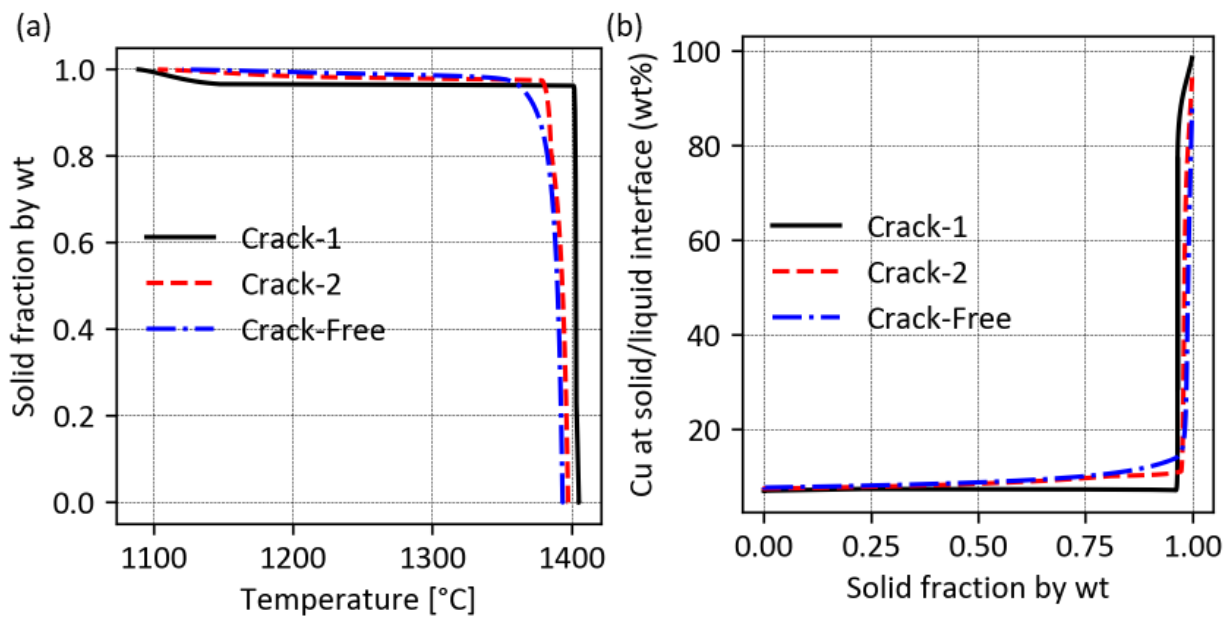

**Figure 1** Scheil simulation results: (a) Solid fraction vs. temperature and (b) Cu concentration at the solid/liquid interface vs. solid fraction curves for three samples with different Ni concentrations show that the curves for the samples with cracking have drastic solidus temperature drops and drastic Cu composition changes at the end of the solidification process while the curves for the crack-free sample show a more gradual reduction in solidus temperature and Cu concentration change. The Ni concentrations in the samples increased gradually from crack-1 (70/30 wt% SS316L/Monel400), to crack-2 (52.5/17.5/30 wt% SS316L/Ni/Monel400), to crack-free samples (38.5/31.5/30 wt% SS316L/Ni/Monel400).

The drastic solidus temperature drop (i.e., the almost horizontal part of the curves for the cracked samples shown in **Figure 1(a)**) and the drastic Cu composition increase at the end of solidification process (i.e., the almost vertical part of the curves for the cracked samples shown in **Figure 1(b)**) was found to be the reason for the hot cracking in the SS316L-Ni-Monel400 three-alloy system. The hybrid Scheil-equilibrium approach developed in ref. [18] was found to successfully detect the above two factors using the formation of a secondary Cu-rich FCC as an indicator of conditions for crack susceptibility. An example is shown in **Figure 2**, whereby a Scheil calculation was first carried out for an alloy composition with 49/29/22 wt% SS316L/Monel400/Ni. Subsequently, equilibrium calculations were performed for compositions along the solidification profile to determine if a transition from Fe-rich FCC to Cu-rich FCC was predicted. Twenty compositions, corresponding to uniform spacing in temperature during solidification, were used for the subsequent equilibrium calculations. In addition, if the difference in Cu concentration between two adjacent selected compositions exceeded 10 wt%, 10 additional compositions were inserted between the points, as illustrated in **Figure 2(a)** to avoid missing the transition from the Fe-rich FCC#1 phase to Cu-rich FCC#2 phase. A temperature of 1080 °C was selected for the equilibrium calculations, which is slightly below the melting temperature of Cu (1084 °C) to avoid liquid being predicted in the equilibrium calculation. The hybrid Scheil-equilibrium method (**Figure 2(b)**) predicted that a secondary Cu-rich FCC will form at the end of the solidification process or later in the HAZ and resulting in hot cracking. **Figure 3** shows the hybrid Scheil-equilibrium results for one of the cracked samples composed of 52.5/17.5/30 wt% SS316L/Ni/Monel400 and the crack-free sample composed of 38.5/31.5/30 wt% SS316L/Ni/Monel400, where the formation of Cu-rich FCC#2 phase is predicted in the cracked sample.

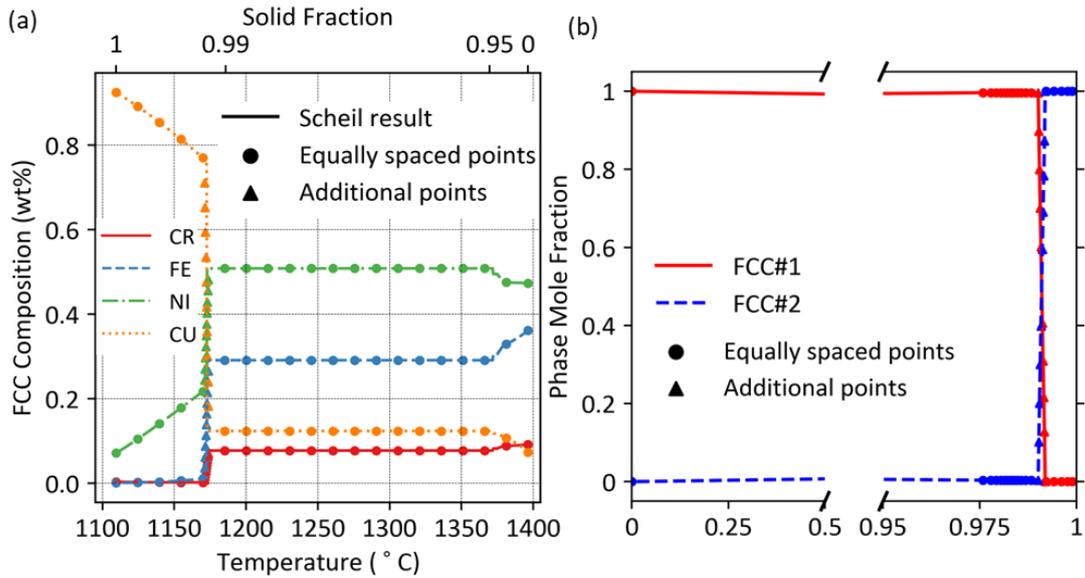

**Figure 2** Results of an example of hybrid Scheil-equilibrium calculations for a composition of 49 wt% SS316L, 29 wt% Monel400, and 22 wt% Ni. This composition lies at the edge of the infeasible region in the SS316L-Ni-Monel400 system. (a) Scheil simulation results with selected compositions for the subsequent equilibrium simulation. Circles represent compositions with equal spacing in terms of temperature, while triangles represent additional compositions inserted when Cu changes significantly over a small temperature range (i.e., larger than 10 wt% difference between two adjacent circle points) to avoid missing the FCC#1 and FCC#2 transition. (b) Hybrid Scheil-equilibrium results at 1080 °C for compositions shown in (a), which predict the transition from Fe- and Ni-rich FCC#1 phase to Cu-rich FCC#2 phase for compositions with a high solid fraction (i.e., in the interdendritic region).

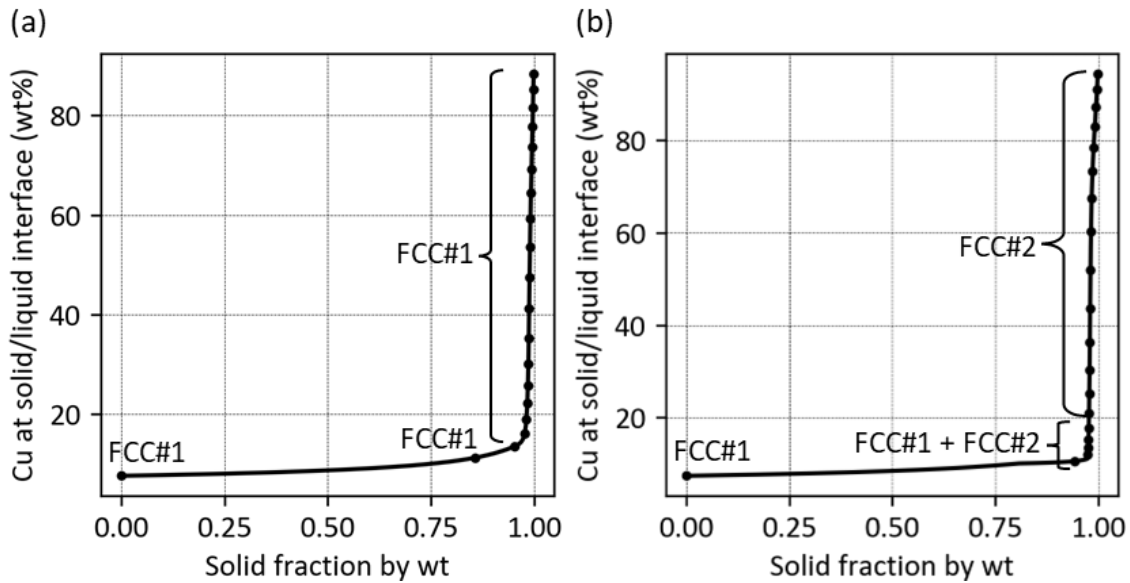

**Figure 3** Hybrid Scheil-equilibrium predicted phases (denoted by the text) at 1080 °C for (a) a crack-free sample composed of 38.5/31.5/30 wt% SS316L/Ni/Monel400 and (b) a sample with cracking composed of 52.5/17.5/ 30 wt% SS316L/Ni/Monel400, which shows only one FCC phase in the crack-free sample and two FCC phases in the sample for which cracking was observed. The symbols represent compositions used for equilibrium simulations, with only Cu concentration shown here for clarity.

## 2.2 Experimental methods

Monolithic multi-layer samples and an FGM joining SS316L to Monel400 through an SS316L+Ni mixing were fabricated by DED AM (LASERTEC 65 DED hybrid, DMG MORI, United States). Samples were deposited on an SS316L substrate when Monel400 amount is lower than 50 wt% and were deposited on an Inconel 625 (IN625) substrate when Monel 400 amount is higher than 50 wt% to avoid the dilution from SS316L substrate resulting in cracking at the sample/substrate interface. **Table 1** shows the alloy compositions, building parameters and cracking results for the monolithic multi-layer samples. Two building parameters were used for these samples because significant porosity problem was observed in the sample with an alloy composition of 40/20/40 wt% SS316L/Monel400/Ni (i.e., sample 8 in **Table 1**, see **Figure 4 (f)**), so building parameters were optimized to reduce porosity for the compositions between 50/50 wt% SS316L/Ni and Monel400 (i.e., sample 9-13 in **Table 1**). An FGM was fabricated from SS316L to Monel 400 through an SS316L+Ni mixture. To address the porosity issue, the SS316L/Ni mixing ratio was adjusted to 55/45 wt% to increase the solidification temperature range (i.e., $T_{liquidus}$-$T_{solidus}$), which increase the time for trapped gas to escape the melting pool. Additionally, only one pass (i.e., one bead for each layer) was designed for the wall sample to reduce residual stress across the melting pool and prevent overlapping regions between the beads, where large pores were commonly observed in monolithic samples. A 3 mm laser beam size was used to fabricate the single-pass 64-layer FGM sample, ensuring a sufficiently thick wall sample for subsequent characterizations. The FGM was fabricated with a laser power of 2500 W, a travel speed of 1000 mm/min, a mass flow rate of 9 g/min, and Ar shroud gas.

Table 1 Alloy compositions, processing parameters and cracking results for the multi-layer monolithic samples fabricated.

| Sample | Alloy Composition | | | Building Parameters | | | | | | Result |
|---|---|---|---|---|---|---|---|---|---|---|
| | SS316L | Monel400 | Ni | Laser Power (W) | Laser Travel Speed (mm/min) | Powder Mass Flow Rate (g/min) | Shroud Gas | Laser Beam Size (mm) | Substrate | Cracking |
| 1 | 100 | 0 | 0 | 900 | 1000 | 6 | None | 1.6 | SS316L | No |
| 2 | 80 | 20 | 0 | 900 | 1000 | 6 | None | 1.6 | SS316L | Yes |
| 3 | 60 | 40 | 0 | 900 | 1000 | 6 | None | 1.6 | SS316L | Yes |
| 4 | 40 | 60 | 0 | 900 | 1000 | 6 | None | 1.6 | IN625 | Yes |
| 5 | 20 | 80 | 0 | 900 | 1000 | 6 | None | 1.6 | IN625 | No |
| 6 | 0 | 100 | 0 | 900 | 1000 | 6 | None | 1.6 | IN625 | No |
| 7 | 52.5 | 30 | 17.5 | 900 | 1000 | 6 | None | 1.6 | SS316L | Yes |
| 8 | 40 | 20 | 40 | 900 | 1000 | 6 | None | 1.6 | SS316L | No |
| 9 | 50 | 0 | 50 | 1100 | 1000 | 4 | Ar | 1.6 | SS316L | No |
| 10 | 40 | 20 | 40 | 1100 | 1000 | 4 | Ar | 1.6 | SS316L | No |
| 11 | 30 | 40 | 30 | 1100 | 1000 | 4 | Ar | 1.6 | SS316L | No |
| 12 | 20 | 60 | 20 | 1100 | 1000 | 4 | Ar | 1.6 | IN625 | No |
| 13 | 10 | 80 | 10 | 1100 | 1000 | 4 | Ar | 1.6 | IN625 | No |

The monolithic multi-layer samples and the FGM sample were sectioned, mounted, and polished using standard metallographic grinding and polishing techniques until a final 0.05 μm finish. Optical microscope (OM) imaging using a digital OM (VHX-2000, Keyence, Japan) was carried out to check for cracking in the as-deposited sample and scanning electron microscope (SEM) imaging, energy-dispersive X-ray spectroscopy (EDS) analysis and electron backscatter diffraction (EBSD) analysis were carried out on an SEM (Apreo 2, ThermoFisher, United States) to characterize composition and grain structure around the cracks in the sample. Vickers hardness measurements were performed (AMH55 hardness tester, LECO, Germany) using a load of 300 gf and a loading time of 15 s to characterize hardness across the as-deposited FGM sample. Five hardness indentations were measured every 500 μm along the sample height on the as-deposited sample.

## 3. Results

### 3.1 Experimentally observed microstructures

Multi-layer samples with a wide range of compositions within the SS316L-Ni-Monel400 system were fabricated as shown in **Figure 4** and **Figure 5**. Cracks were observed in monolithic 20 wt%, 40 wt%, and 60 wt% Monel 400 (balance SS316L) (see **Figure 4(a-c)**), while no cracks were observed in monolithic 80 wt% Monel400 (balance SS316L) (see **Figure 4(d)**). SEM images and EDS maps for the sample with an alloy composition of 40/60 wt% SS316L/Monel400, in which cracks were observed, are shown in **Figure 6**; Cu microsegregation is observed near the crack. The Cu microsegregation shown in **Figure 6(b)** indicates the cracks in the 40/60 wt% SS316L/Monel400 samples are solidification cracks due to the existence of a low melting point Cu-rich liquid film in interdendritic regions at the end of the solidification process.

A 10-pass (i.e., 10 beads within each layer) 10-layer 52.5/30/17.5 wt% SS316L/Monel400/Ni FGM sample was deposited onto an SS316L substrate to evaluate if increasing Ni concentration would sufficiently reduce Cu microsegregation to avoid hot cracking. Fewer and less severe cracks were observed in this sample as shown in **Figure 4(e)**.

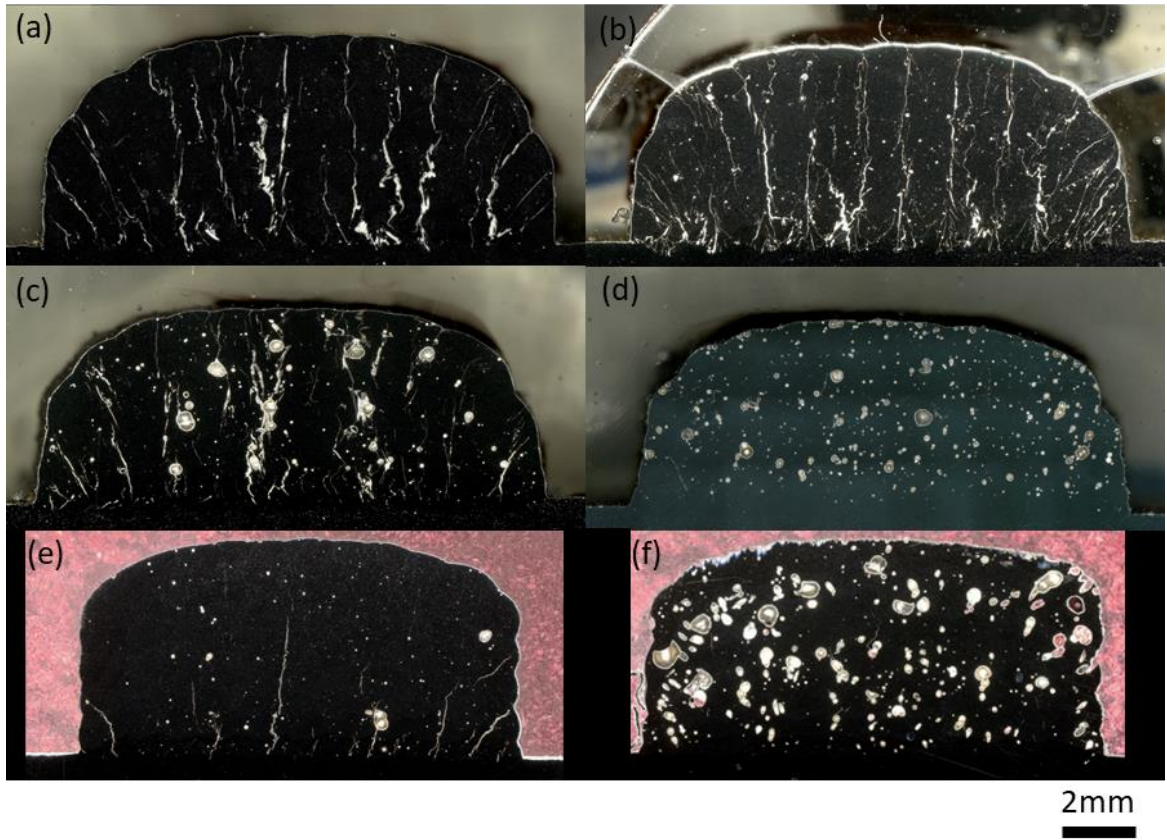

**Figure 4** OM images of the 10-pass 10-layer monolithic samples fabricated using 900 W laser power, 1000 mm/min laser travel speed, and 6 g/min powder mass flow rate: (a) sample with 80/20 wt% SS316L/Monel400 on SS316L substrate; (b) sample with 60/40 wt% SS316L/Monel400 on SS316L substrate; (c) Sample with 40/60 wt% SS316L/Monel400 on IN625 substrate; (d) sample with 20/80 wt% SS316L/Monel400 on IN625 substrate; (e) sample with 52.5/30/17.5 wt% SS316L/Monel400/Ni on SS316L substrate and (f) sample with 40/20/40 wt% SS316L/Monel400/Ni on SS316L substrate.

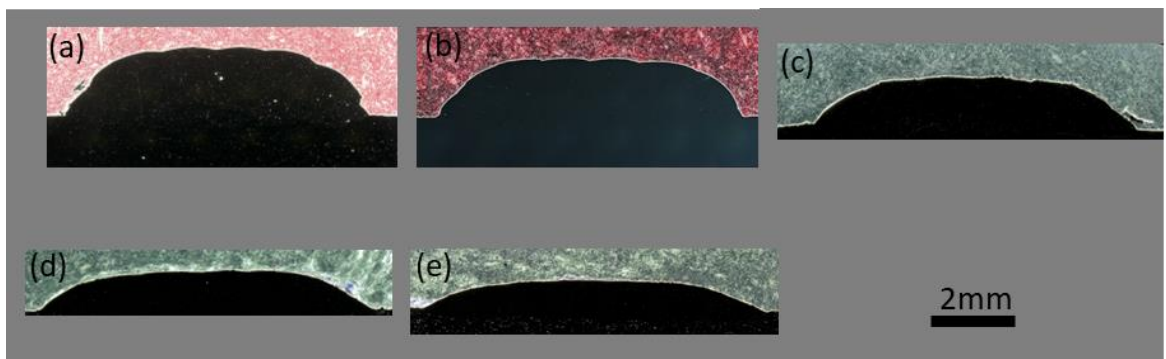

**Figure 5** OM images of monolithic samples fabricated using 1100 W laser power, 1000 mm/min laser travel speed and 4 g/min powder mass flow rate. (a-e) 5-pass 5-layer samples with composition from 50/50 wt% SS316L/Ni to Monel400: (a) sample with 50/50 wt% SS316L/Ni on SS316L substrate; (b) sample with 40/20/40 wt% SS316L/Monel400/Ni on SS316L substrate; (c) sample with 30/40/30 wt% SS316L/Monel400/Ni on SS316L substrate; (d) sample with 20/60/20 wt% SS316L/Monel400/Ni on IN625 substrate; and (e) sample with 10/80/10 wt% SS316L/Monel400/Ni on IN625 substrate.

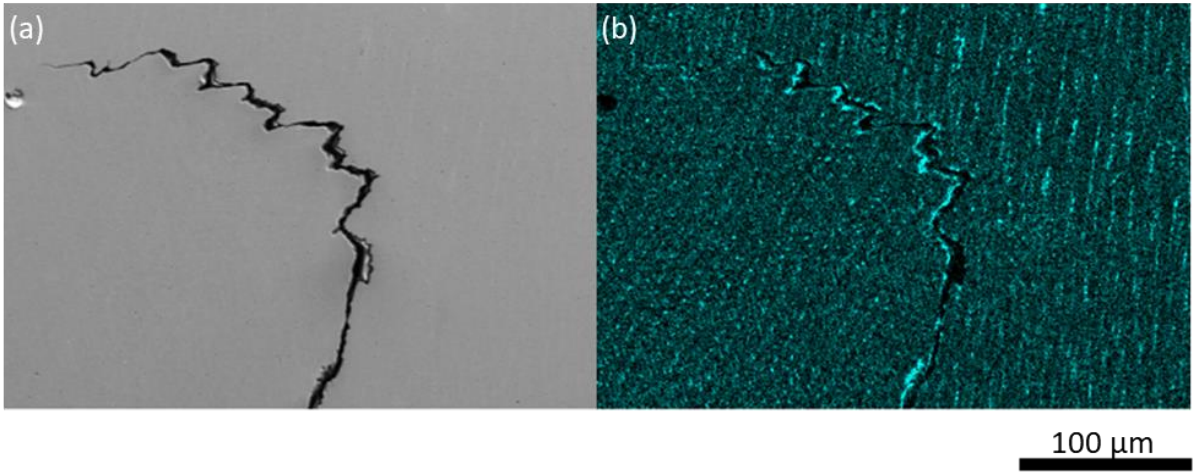

**Figure 6** SEM image (a) and EDS maps of Cu (b) of a solidification crack in a monolithic 10-layer samples with 40/60 wt% SS316L/Monel400 showing Cu microsegregation near the crack, suggesting that the crack is formed due to hot cracking.

### 3.2 Feasibility map for SS316L-Ni-Monel400 system

The feasibility map developed solely for avoiding deleterious phases in ref. [10] predicts only FCC and BCC phases, thus no brittle intermetallic phases, in the SS316L-Ni-Monel400 system. To consider the propensity for cracking, the improved crack susceptibility criterion (iCSC) heat map (ref. [18], [19]) was calculated for the SS316L-Ni-Monel400 system as shown in **Figure 7,** accurately predicting cracking in most of the samples in which cracking was observed except for the 80/20 wt% SS316L/Monel400 sample and the 52.5/30/17.5 wt% SS316L/Monel400/Ni sample where cracking was observed but was not predicted.

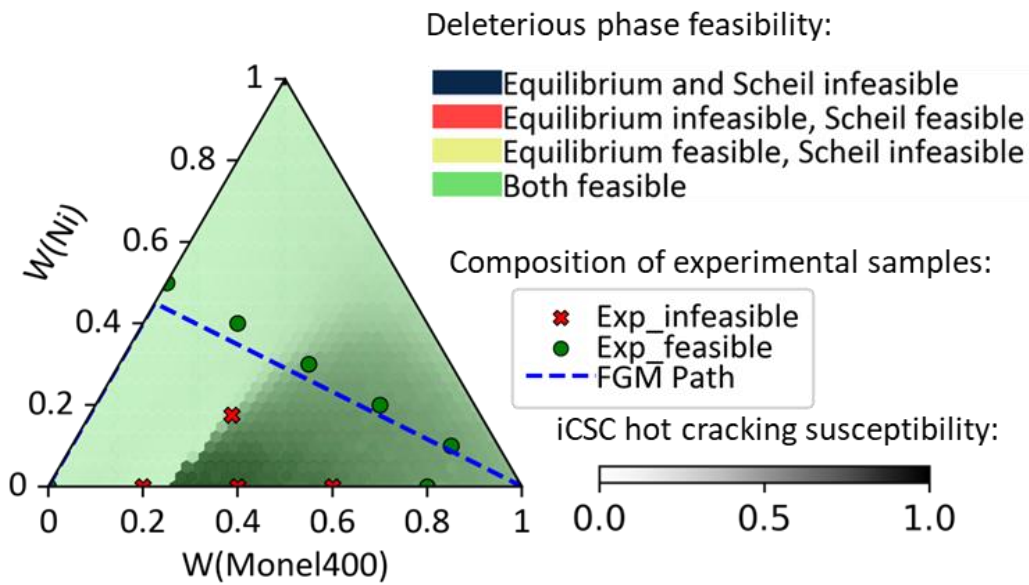

**Figure 7** Deleterious phase feasibility map and crack susceptibility heat map based on the iCSC criterion calculated using the method in ref. [10] and [18] for the SS316L-Ni-Monel400 three-alloy system, simplified as the Cr-Cu-Fe-Ni system. Symbols in the map indicate compositions for which experiments were performed, with the color and shape of the symbol corresponding to whether cracks were (red 'x') or were not (green circle) experimentally observed.

In all cases here, cracking resulted from significant Cu microsegregation within the SS316L-Ni-Monel400 system. For example, despite the overall composition of the 52.5/30/17.5 wt% SS316L/Monel400/Ni sample containing only approximately 10 wt% Cu, Scheil calculation for this composition predicted Cu concentration in the last-solidified FCC to be approximately 95 wt% (see **Figure 2(b)**). The formation of Cu-rich liquid film (see the Cu microsegregation in **Figure 2(b)**) and the corresponding drop in solidification temperature (see **Figure 2(a)**) occur at the very end of the solidification process (solid >97.5 wt%). The effect of Cu microsegregation at the very end of the solidification process is usually underestimated in the hot cracking susceptibility criteria because they consider a wider solid fraction range (e.g., 0.7-0.98 for iCSC criterion). However, due to the low wetting angle between the Cu-rich liquid and the Fe-rich matrix [5], Cu-rich liquid can infiltrate easily between solidified microstructural features (e.g., dendrites and grains) resulting in a continuous liquid film forming even with a small amount of remaining liquid, which can result in hot cracking.

The problem cannot be solved by simply adjusting the solid fraction range considered in the hot cracking susceptibility criteria because these consider only the solid fraction vs. temperature curve during the solidification process, and the solidus temperature may drop rapidly at the end of solidification for crack-free compositions as well (see the curve for the crack-free composition in **Figure 2(a)**). The hot cracking susceptibility criteria also overestimated the hot cracking susceptibility for compositions with high Ni concentration. **Figure 7** shows that relatively high hot cracking susceptibility was predicted for the 20/80 wt% SS316L/Monel400 and 100 wt% Monel400 samples, despite no observed cracks in experiments (see **Figure 4(d)**). This occurs because the solidification temperature decreases rapidly, attributed to Cu microsegregation throughout the solidification process for these compositions. These temperature drops were captured by the iCSC criterion and resulted in a high hot cracking susceptibility prediction. However, with a higher Ni concentration in both the matrix and remaining liquid for these compositions, the decrease in solidus temperature as well as Cu microsegregation during solidification is more gradual (see **Figure 2**) than those for the compositions of the cracked samples and thus continuous liquid film is unlikely to form.

As discussed in section 2.1, the hybrid Scheil-equilibrium approach was used to predict the formation of Cu-rich FCC in interdendritic regions as an indicator of the drastic solidus temperature drop and Cu microsegregation at the end of the solidification process for the compositions of the cracked samples (see **Figure 2**). The prediction of a secondary Cu-rich FCC phase indicates that there is significant Cu microsegregation in the interdendritic region, resulting in a significant composition difference between the interdendritic Cu-rich FCC#2 phase and the matrix Fe-rich FCC#1 phase (see **Figure 3**) and the formation of continuous liquid film and thus hot cracking. For compositions with high Ni concentrations, although Cu will still microsegregate, no transition from Fe-rich FCC#1 phase to Cu-rich FCC#2 phase is predicted along the solidification compositions due to high Ni concentration and gradual composition change during the solidification, thus hot cracking is not predicted.

The feasibility map for Cu-microsegregation-induced hot cracking for the SS316L-Ni-Monel400 calculated using an optimized Cr-Cu-Fe-Ni database (ref. [20]). This database considers all binary (Cr-Cu [21], Cr-Fe, Cr-Ni, Cu-Fe [22], Cu-Ni [21] , Fe-Ni) and ternary systems(Cr-Cu-Fe, Cr-Cu-Ni [21], Cr-Fe-Ni, Cu-Fe-Ni). Some existing systems were collected from literature and the missing ones were developed. The feasibility map is given in **Figure 8** showing that when the Cu concentration increases near the SS316L corner, corresponding to when Monel400 is added, hot cracking due to Cu microsegregation is predicted. However, when additional Monel400 or Ni is added, the decrease in Fe and increase in Ni reduces the Cu microsegregation and thus the propensity for hot cracking.

No cracking was seen in the monolithic sample with 40/20/40 wt% SS316L/Monel400/Ni (sample 8 in **Table 1**), suggesting that increasing Ni concentration avoided hot cracking, but large pores were observed (see **Figure 4(f)**). To reduce the porosity in the remaining samples, a lower powder mass flow rate and higher laser power were used for samples 9-13 in **Table 1** so that the bead will be flatter and wider, reducing the travel distance for trapped gas to escape the melting pool. The samples with compositions from 50/50 wt% SS316L/Ni to pure Monel400 were fabricated using the new building parameter (see **Figure 5**) and validated the lack of cracking predicted computationally.

The compositions of the monolithic multi-layer samples are plotted on the feasibility map calculated based on the new hybrid Scheil-equilibrium method (see **Figure 8**), where red "x" symbols represent samples in which cracking was observed and green circles represent crack-free samples. All experimental results match the designated feasible and infeasible regions predicted in the feasibility map.

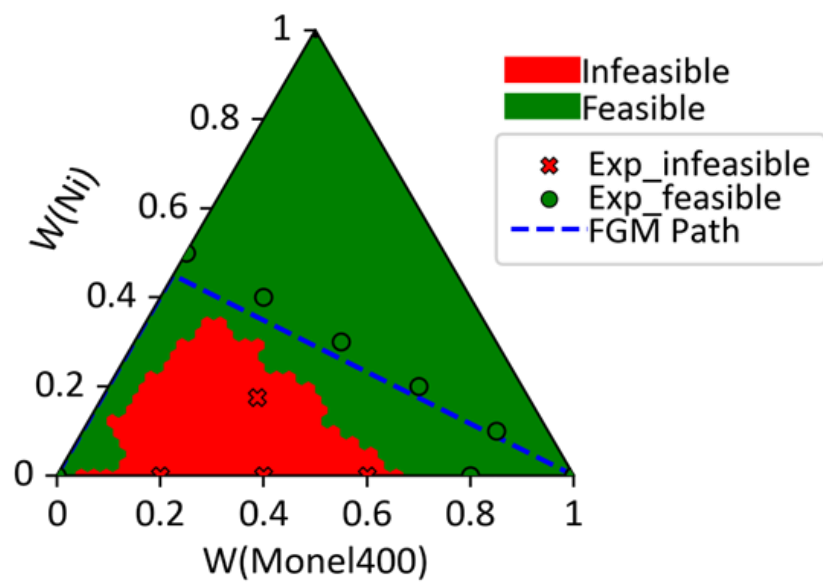

**Figure 8** Feasibility map based on the new hot cracking criterion (i.e., whether a secondary Cu-rich FCC phase was predicted by the hybrid Scheil-equilibrium approach) for the SS316L-Ni-Monel400 three-alloy system (simplified as Cr-Cu-Fe-Ni). Symbols represent experimentally probed compositions.

## 3.3 SS316L-SS316L/Ni-Monel400 FGM

An FGM grading vertically from SS316L to 55/45 wt% SS316L/Ni to Monel400 was fabricated on an SS316L substrate. A total of 7 layers of 55/45 wt% SS316L/Ni, 7 layers of 27.5/22.5/50 wt% SS316L/Ni/Monel400, and 50 layers of Monel400, were deposited to achieving a final FGM sample with a height of approximately 20 mm and length of 150 mm (see **Figure 9(a)**). EDS line scans along the centerline of the sample (see **Figure 10(a)**) confirmed that the designed composition was successfully deposited. The spatial distribution of Cu in the gradient region of the FGM sample, at a layerwise composition of 32/26/42 wt% SS316L/Ni/Monel400, is shown by the EDS image in **Figure 11**, where the Cu microsegregation was significantly reduced compared to that in the cracking samples (see **Figure 6**).

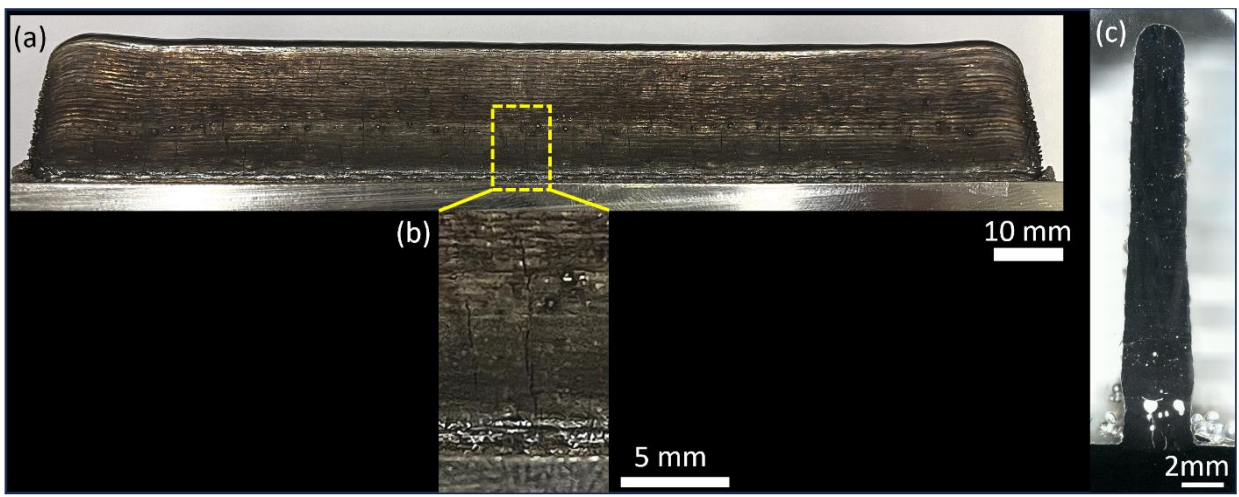

**Figure 9** Photographs of the single-pass multi-layer SS316L to 55/45 wt% SS316L/Ni to Monel400 FGM sample: (a) view of the side of the FGM wall; (b) zoomed-in image of vertical cracks at the center of the wall; and (c) OM image of the FGM cross-section.

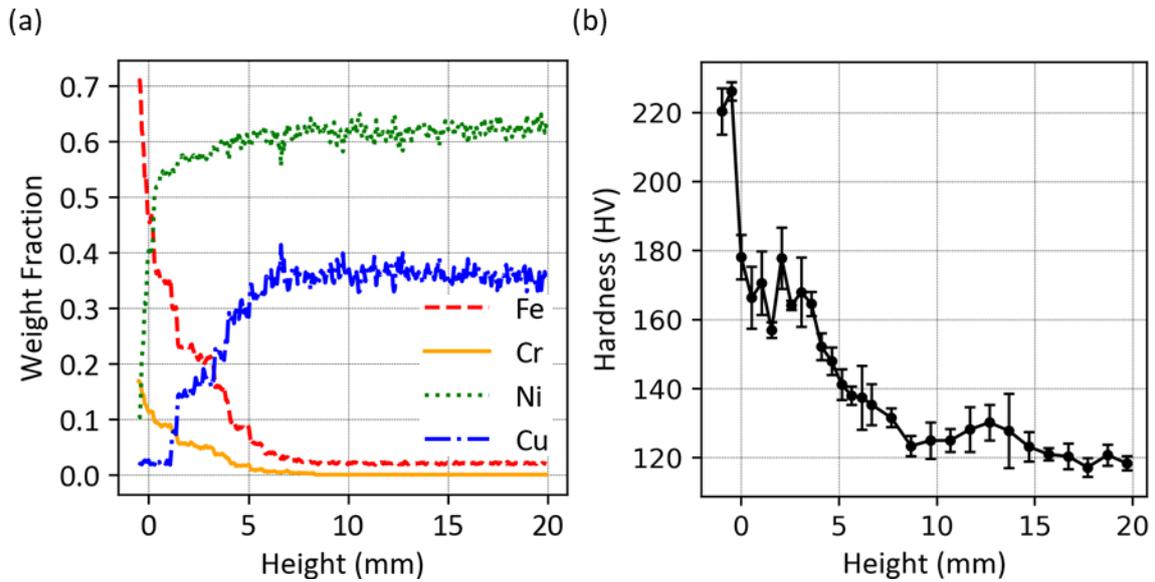

**Figure 10** (a) Composition along the height of the FGM measured using EDS line scans and (b) hardness along the height of the FGM.

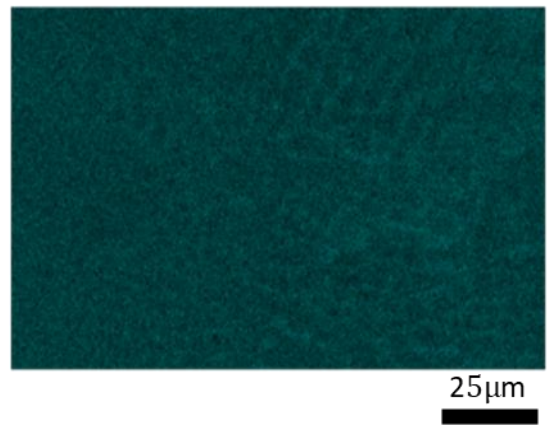

**Figure 11** Cu EDS map for the region with ~32/26/42 wt% SS316L/Ni/Monel400 in the FGM sample showing a reduction in Cu microsegregation compared to that in the SS316L-Monel400 samples shown in **Figure 6**.

Pores were found at the bottom of the FGM sample in the OM image in **Figure 9(c)**. These pores were located primarily at 33/27/40 wt% SS316L/Ni/Monel400 and are likely due to gas trapped in the melting pool during fabrication due to the small solidification temperature range for compositions with low Monel400 concentrations, resulting in insufficient time for the trapped gas to escape. It is hypothesized that these pores could be removed with sample heating, reducing the layer height through the reduction of the mass flow rate of powder, or by reducing laser power and scan speed to stabilize the liquid in the melt pool in future fabrications.

Vertical cracks were observed in the center of the FGM as shown in **Figure 9(b)**. As these cracks were mostly vertical and observed at the middle of the FGM sample, at least 2 cm away from the ends of the wall, and also found in the Monel400 region of the FGM sample, it is hypothesized that these cracks are due to horizontal residual stresses resulting from depositing a long single-pass wall sample rather than being driven by compositional issues. This is supported by the lack of Cu microsegregation in the gradient region of the FGM sample, as shown in **Figure 11**, the predominantly vertical orientation of the cracks located in the middle of the FGM wall sample (see **Figure 9 (b)**), and the presence of cracks in the Monel400 region, which is expected to be crack-free based on its composition. Thus, these observed cracks should be able to be removed by fabricating shorter samples in the future, or modifying the thermal environment with heating, to reduce residual stress.

The Vickers microhardness decreased from about 220 HV in the SS316L substrate to about 170 HV in the gradient region, ranging from 55/45 wt% SS316L/Ni to 27.5/22.5/50 wt% SS316L/Ni/Monel400, and further decreased to about 120 HV in the Monel400 region (see **Figure 10(b)**). The variance of the hardness in the 55/45 wt% SS316L/Ni and the 27.5/22.5/50 wt% SS316L/Ni/Monel400 region is most likely a result of pores beneath the sample surface that impact hardness measurements.

The designed compositional path for the SS316L-55/45 wt% SS316L/Ni-Monel400 FGM sample was also plotted in the feasibility map shown in **Figure 8**, matching the feasible region predicted by the hybrid Scheil-equilibrium method. The FGM sample, together with the monolithic multi-layer samples, demonstrated the feasibility of avoiding hot cracking between SS316L and Monel400 by modifying Ni concentration to reduce Cu microsegregation in the gradient region along with the capability of using the hybrid Scheil-equilibrium method to predict hot cracking in the SS316L-Ni-Monel400 three-alloy system.

## 4. Conclusions

In the present work, cracking in the SS316L-Ni-Monel400 three-alloy system was studied. A new cracking criterion for joining Fe-alloy and Cu-containing alloy was developed based on the hybrid Scheil-equilibrium method and validated by monolithic multi-layer samples and an FGM sample in the SS316L-Ni-Monel400 three-alloy system. The new hot cracking criterion, based on the hybrid Scheil-equilibrium approach, is expected to be applicable not only to the SS316L-Ni-Monel400 system but also to other Ni and Cu alloys such as Inconel or bronze.

The key findings of this study were as follows:

- The low solubility between Cu and Fe results in significant Cu microsegregation during the solidification process, leading to solidification cracking or liquation cracking when joining the SS316L and Monel400. Modifying the Ni concentration in the gradient region between SS316L and Monel400 was found to successfully eliminate hot cracking.

- The previously developed feasibility map for deleterious phases focused on cracking under solidus temperature due to brittle intermetallic phases. The hot cracking susceptibility map successfully predicted cracking in most monolithic multi-layer samples except for the cracking in the monolithic multi-layer sample with 80/20 wt% SS316L/Monel400. A new hot cracking criterion was developed based on whether a Cu-rich secondary FCC phase was predicted by the hybrid Scheil-equilibrium approach. It successfully predicted the cracking region in the SS316L-Ni-Monel400 three-alloy system and was validated by monolithic multi-layer and FGM samples. Compared to the other hot cracking susceptibility criteria, the new approach captures the Cu microsegregation at the very end of the solidification process and differentiates the Cu microsegregation between the low-Ni-concentration infeasible (i.e., cracking) compositions and the high-Ni-concentration feasible (i.e., crack-free) compositions. The new hot cracking criterion also has potential use for other Fe-alloy, Ni-alloy, and

- Cu-containing alloy systems.
- Monolithic multi-layer samples with a wide range of composition in the SS316L-Ni-Monel400 system and an FGM from SS316L to Monel 400 with modified Ni concentrations in the gradient region were successfully fabricated. The cracks in the FGM sample were found to be a result of residual stress instead of a composition issue and the pores in the FGM sample were a result of trapped gas. Both defects can be avoided by adjusting sample geometry and building parameters.

**Acknowledgements:**

The authors are grateful for financial support from the National Science Foundation (NSF Grant CMMI-2050069) and the Office of Naval Research (ONR Grant N00014-21-1-2608). The authors are also grateful for Arin Kim Lee and Sarah Lynne Bury for their help with sample preparation in this study and for Cory Jamieson and Christopher Schaffer for their help with sample fabrication.

**Declaration of competing interest:**

None

**Data availability**

All relevant data is available from the authors.